\documentclass[a4paper,twocolumn,english,showpacs,aps,prl,fleqn,superscriptaddress]{revtex4-1}
\usepackage[T1]{fontenc}
\usepackage[utf8]{luainputenc}
\usepackage{amsmath}
\usepackage{amssymb}
\usepackage{graphicx}
\usepackage{hyperref}
\usepackage{babel}
\usepackage{wasysym}
\usepackage{microtype}
\usepackage{color}
\usepackage{bm}

\makeatletter
\usepackage{indentfirst}
\usepackage{bm}
\usepackage{graphicx}
\def\@hangfrom@section#1#2#3{\@hangfrom{#1#2#3}}     

\renewcommand\section{\@startsection{section}{1}{\z@}
{3.25ex \@plus1ex \@minus.2ex}
{1.5ex \@plus.2ex}
{\normalfont\normalsize\bfseries}}   

\renewcommand\subsection{\@startsection{subsection}{2}{\z@}
{3.25ex \@plus1ex \@minus.2ex}
{1.5ex \@plus.2ex}
{\normalfont\normalsize\bfseries}}

\makeatother

\begin{document}
\renewcommand\figurename{FIG.}
\setlength{\parskip}{0.3em}
\title{Oscillatory-like Expansion of a Fermionic Superfluid}

\author{Xiao-Qiong Wang}
\author{Yu-Ping Wu}
\author{Xiang-Pei Liu}
\author{Yu-Xuan Wang}
\author{Hao-Ze Chen}
\author{Mudassar Maraj}
\author{Youjin Deng}
\affiliation{Shanghai Branch, National Laboratory for Physical Sciences at Microscale and Department of Modern Physics, University of Science and Technology of China, Shanghai, 201315, China}
\affiliation{CAS Center for Excellence and Synergetic Innovation Center in Quantum Information and Quantum Physics, University of Science and Technology of China, Hefei, Anhui 230026, China}

\author{Xing-Can Yao}
\email{yaoxing@ustc.edu.cn}
\author{Yu-Ao Chen}
\email{yuaochen@ustc.edu.cn}
\author{Jian-Wei Pan}
\email{pan@ustc.edu.cn}

\affiliation{Shanghai Branch, National Laboratory for Physical Sciences at Microscale and Department of Modern Physics, University of Science and Technology of China, Shanghai, 201315, China}
\affiliation{CAS Center for Excellence and Synergetic Innovation Center in Quantum Information and Quantum Physics, University of Science and Technology of China, Hefei, Anhui 230026, China}

\begin{abstract}
We study the expansion behaviours of a Fermionic superfluid in a cigar-shaped optical dipole trap for the whole BEC-BCS crossover and various temperatures. At low temperature ($0.06(1) T_F$), the atom cloud undergoes an anisotropic hydrodynamic expansion over 30~ms, which behaves like oscillation in the horizontal plane. By analyzing the expansion dynamics according to the superfluid hydrodynamic equation, the effective polytropic index $\overline{\gamma}$ of Equation-of-State of Fermionic superfluid is extracted. The $\overline{\gamma}$ values show a non-monotonic behavior over the BEC-BCS crossover, and have a good agreement with the theoretical results in the unitarity and BEC side. The normalized quasi-frequencies of the oscillatory expansion are measured, which drop significantly from the BEC side to the BCS side and reach a minimum value of 1.73 around $1/k_Fa=-0.25$. Our work improves the understanding of the dynamic properties of strongly interacting Fermi gas.

\end{abstract}

\maketitle

\section{Introduction}

With an exquisite control of interaction strength and other physical parameters~\cite{Chin2010RMP,Dieckmann2002PRL},
strongly interacting Fermi gas has become a versatile platform to study a wide variety of fundamental questions and phenomena,
ranging from high-temperature superconductor, neutron star to quark-gluon plasma of the early Universe~\cite{Bloch2008RMP,Stringari2008RMP}.
Important experimental progresses include realization of molecular Bose-Einstein condensate (BEC)~\cite{Greiner2003,Jochim2003},
generation of vortices in a rotating Fermi gas as a conclusive evidence of Fermionic superfluidity~\cite{Zwierlein2005,Zwierlein2006Nature},
and precise study of superfluid phase transition and universal properties via direct measurement of  the Equation-of-State (EoS) of Fermi gas~\cite{Nascimbene2010Nature,Navon2010Nature,Horikoshi2010,Ku2012Nature} etc.

\begin{figure*}[htbp]
  \centering
  \includegraphics[width=0.85\textwidth]{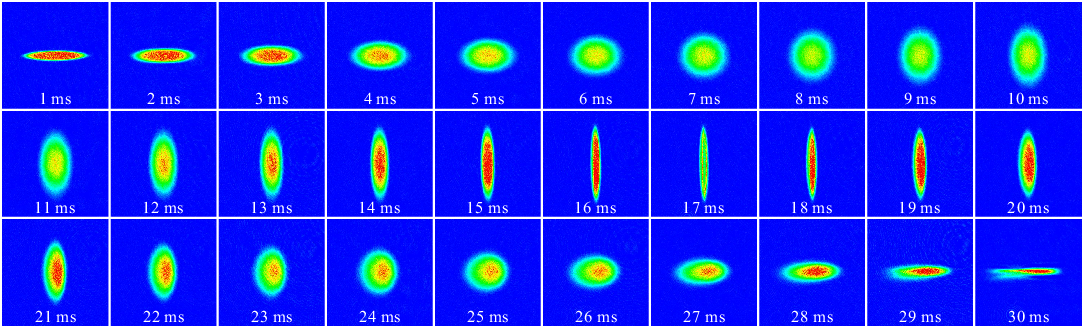}
  \caption{Oscillatory-like expansion in the $xy$ plane of $^6$Li superfluid at 832 G. Each picture has a field of view of $1.3$~mm$\times$$1.3$~mm.}\label{figure1}
\end{figure*}

Among various techniques for gaining information on the properties of strongly interacting Fermi gas, measurement of the expansion dynamics is established to be a powerful tool, i.e., by suddenly switching off the strong confining optical dipole trap and letting the Fermi gases expand in the residual magnetic curvature potential~\cite{OHara2002,ketterle2008making,Trenkwalder2011PRL}. It provides valuable information about the state of Fermi gas and the role of interactions. For example, by analyzing the expansion dynamics, three different regimes can be distinguished, i.e., the collisionless~\cite{Jochim2014PRA}, classical hydrodynamic~\cite{Bourdel2003PRL}, and superfluid hydrodynamic expansions. In the past decades, great efforts have been devoted to the study of expansion dynamics of Fermi gas. Based on a collisionless ballistic expansion, the initial momentum distribution of strongly interacting Fermi gas can be precisely probed through matter-wave focusing technique~\cite{Jochim2014PRA,Jochim2019Science}. Collisional hydrodynamic behaviors~\cite{OHara2002} have been well revealed through the observation of a $2$~ms anisotropic expansion of the strongly interacting Fermi gas. The interaction energy of Fermi gas near the Feshbach resonance has been measured through the study of short-time ballistic (anisotropic) expansions at weakly (strongly) interacting regimes~\cite{Bourdel2003PRL}. Hydrodynamic expansion of a strongly interacting Fermi-Fermi mixture up to 8 ms has been studied, leading to the observation of anisotropic expansion and hydrodynamic drag between the two species~\cite{Trenkwalder2011PRL}. Very recently, shear viscosity~\cite{Cao2011NJP,Elliott2014PRLb,Joseph2015PRL}, conformal symmetry breaking and scale invariance~\cite{Elliott2014PRL} have been explored through the measurement of expansion dynamics. However, while being intensively studied, the observation of oscillatory-like expansion of superfluid hydrodynamics, which simultaneously requires large atom number, low temperature, and special tailored experimental configuration, has not been reported.

In this Letter, we prepare a Fermionic superfluid of about $5\times 10^6$ $^6$Li atoms at temperature $T/T_F=0.06(1)$ in a cigar-shaped optical dipole trap,
with $T_F$ being the Fermi temperature. We suddenly switch off the optical trap and observe an anisotropic expansion of the superfluid in a weak residual magnetic field curvature with attractive (repulsive) potential in the horizontal $xy$ plane (gravity $z$ direction), persisting for a long time of more than $30$~ms. A pronounced feature is the out-of-phase oscillatory-like expansion behavior in the horizontal plane; namely, the cloud size expands (shrinks) in the radial direction while shrinks (expands) in the axial direction during the expansion. This phenomenon is a manifestation of hydrodynamics, since for a low-temperature collisionless gas, the oscillatory-like expansion should be in-phase, i.e., the cloud size expands or shrinks simultaneously in the radial and axial directions. We further show that the expansion dynamics is qualitatively consistent with the numerical solution of a hydrodynamic equation of motion for trapped Fermionic superfluid and quantitative agreement is observed during the first several milliseconds of expansion.

From a least-squares-criterion fitting, we determine the effective polytropic index $\overline{\gamma}$ of EoS of the trapped Fermionic superfluid. In the whole region of BEC-BCS crossover, the $\overline{\gamma}$ values display a non-monotonic behavior as a function of interaction strength~\cite{HuiHu2004PRL,ketterle2008making}, and reduce to the well-known theoretical results in the BEC, BCS and unitary limits.  The trap averaged polytropic index $\gamma'$ are also calculated based on the measurements of EoS in ref.~\cite{Navon2010Nature} and a simple weight function~\cite{HuiHu2004PRL}, and have a good agreement with the experimental $\overline{\gamma}$ values in the BEC side. The quasi-frequencies of the expansion are further studied, which show a strong dependence on the interaction strength and the temperature of the cloud. At the lowest temperature, the quasi-frequencies have a similar trend with the numerical results using $\overline{\gamma}$ and $\gamma'$, respectively. As the temperature goes up, interaction strength dependence of the quasi-frequency is smoothed out, mainly due to the decrease of superfluid component. The long-lived oscillatory expansion dynamics and the successful derivation of quantitative information can be attributed to the large atom number and the sufficiently low temperature of the prepared Fermionic superfluid. The $\overline{\gamma}$ values allow us to characterize the dynamic behaviors of the harmonic trapped strongly interacting Fermi gas using hydrodynamic equation,  such as collective oscillations~\cite{Altmeyer2007}, sound modes, and dark/bright solitons~\cite{wenwenPRB2010}.

\section{Experiment}

The experimental setup has been described in our previous works{~\cite{Pan2017,XingCanYao2016,Chen2016PRA,Chen2016}}.
We start from a two-species magneto-optical trap of $^6$Li and $^{41}$K atoms.  To achieve low temperature and high phase space density, D1 line gray molasses for $^{41}$K atoms and UV MOT for $^6$Li atoms are implemented, respectively. Then, both  species are transferred to the magnetic quadrupole trap and adiabatically transported to a full-glass science cell with extremely high vacuum environment, where an optically plugged magnetic trap is used to confine the atom cloud. Next, forced radio-frequency (RF) evaporative cooling is applied on $^{41}$K atoms, while the $^6$Li atoms are cooled sympathetically. After the final radio-frequency (rf) "knife" of evaporation~\cite{Ketterle1996181}, the $^{41}$K atoms are exhausted and a pure $^6$Li cloud is prepared. Then about $2\times 10^7$ $^6$Li atoms with a temperature of about 50~$\mu$K are loaded into the cigar-shaped optical dipole trap (wavelength 1064~nm, $1/e^2$ beam waist 64~$\mu$m) and immediately transferred to the lowest hyperfine state by a $3$~ms Landau-Zener sweep~\cite{Shytov2004PRA}.
Next, we adiabatically ramp the magnetic field $B$ to the unitary point at $B=832$~G~\cite{PhysRevLett.110.135301},  and apply several rf pulses to prepare a balanced mixture of the two lowest hyperfine states.
Final evaporative cooling is accomplished by exponentially reducing the depth of the dipole trap.
After a $3$~s evaporation, about $5\times10^6$ $^6$Li atoms at temperature $ T/T_F=0.06(1)$ are prepared.
The radial confinement is mainly optical with  final trap frequencies of $\omega_y=2\pi\times 211.9$~Hz and $\omega_z=2\pi \times 200.1$~Hz,
where $y$ is the radial direction in the horizontal plane and $z$ is the gravity direction.
The axial confinement ($x$ direction) is mainly provided by the magnetic field curvature with a trap frequency $\omega_x= 2\pi\times 16.5$~Hz at 832 G.

\begin{figure*}[htbp]
  \centering
  \includegraphics[width=0.9\textwidth]{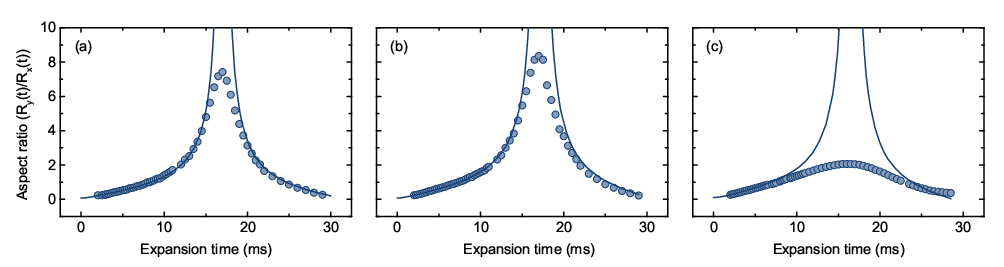}
  \caption{Aspect ratio versus expansion time. Blue circles in (a), (b), and (c) are experiment results taken at 767 G, 832 G, and 1003.2 G, respectively, while blue solid lines are theoretical predictions of Eq.~\ref{equ1} with $\bar{\gamma}=1$ (a), $\bar{\gamma}=2/3$ (b), and $\bar{\gamma}=2/3$ (c), respectively.  Each data point represents the average of four measurements with the error bar (if larger than the symbol) given by the standard deviation. }\label{figure2}
\end{figure*}

Before expansion, the $^6$Li superfluid cloud is held for another $400$~ms to achieve fully thermal equilibrium.
Then the optical dipole trap is abruptly switched off and the $^6$Li cloud is expanded in the residual magnetic field curvature. To minimize the asymmetric effect caused by the repulsive potential in $z$ direction, a magnetic field gradient of $\partial_zB=1.1$~G/cm is simultaneously turned on to levitate the superfluid at the saddle point. An imaging system with resolution of $2.2$~$\mu$m is employed to obtain high quality images~\cite{XingCanYao2016}. The experimental results are shown in Fig.~\ref{figure1}, where pictures are taken at a time interval of 1~ms.
Thanks to the large atom number of the prepared $^6$Li superfluid,  the hydrodynamic expansion persists for a long time of over $30$~ms. Particularly, the full refocusing of atom cloud after the inversion of the aspect ratio is remarkable. It's seen that the cloud sizes in the radial and axial directions both oscillate as a function of expansion time $t$ in the $xy$ plane. We thus define the aspect ratio of the cloud as $R_y(t)/R_x(t)$, where $R_x(t)$ and $R_y(t)$ are respectively the radii in the $x$ and $y$ directions. With an initial value of $0.078$ at $t=0$, the aspect ratio increases, reaches a maximum value of $8.5$ at time $t_0 \sim 17$ ms, and gradually decreases after $t_0$, as shown in Fig.~\ref{figure2} (b). We further define quantity ${f_0}/{f_x} \equiv \frac{1}{2 t_0}/\frac{\omega_x}{2\pi}$ as normalized quasi-frequency and  $A_0\equiv \frac{R_y(t_0)}{R_x(t_0)}/\frac{R_x(0)}{R_y(0)}$ as the normalized quasi-amplitude of the oscillatory expansion, where $\omega_x$ is magnetic field B dependent residual frequency. The obtained ${f_0}/{f_x}$ is 1.806(2), which is smaller than the well known value of 2 for the collisionless gas. Therefore, as anisotropic expansion observed in ref.~\cite {OHara2002}, the realized oscillatory-like expansion is a beautiful illustration of hydrodynamic behavior of strongly interacting Fermi gas.

\section{Quantitative analysis}

The polytropic form of EoS, $\mu(n)\propto n^{\gamma}$, relating the local chemical potential $\mu$ to the density n, provides a valid and convenient description for the strongly interacting Fermi gases in the BEC-BCS crossover, and has been widely used in the analysis of collective dynamics and thermodynamics therein~\cite{ketterle2008making,Menotti2002PRL,Heiselberg2004PRL,Stringari2005PRL,HuiHu2004PRL}. At zero temperature, the polytropic index $\gamma$ depends solely on the interaction strength $1/k_Fa$ of the system. By neglecting the correlations between the dimers, $\gamma$ in the BEC-BCS crossover has been given by the mean-field theory. To further address the strongly interaction effect, quantum Monte Carlo simulations have been carried out to obtain a more precise $\gamma$ along the BEC-BCS crossover~\cite{Giorgini2004PRL,Manini2005PRA}. Very recently, the zero temperature EoS in the strongly interacting regime has been measured by analyzing the density distributions of the Fermi gas~\cite{Ku2012Nature}. The $\gamma$ has been directly derived, agreeing well with the quantum Monte Carlo simulation in the BEC-BCS crossover. However, in most of the experiments, the superfluid is trapped in a harmonic potential, thus a trap-dependent effective polytropic index $\bar{\gamma}$ becomes the key parameter to model the system, which has not been measured previously.

Based on the $\bar{\gamma}$, the superfluid hydrodynamic equation at zero temperature can be written as:
\begin{equation}
\label{equ1}
  \ddot{b_i}(t)=-\omega_i^2(t>0)b_i(t)+\frac{\omega_i^2(t<0)}{b_i(t) V(t)^{\bar{\gamma}}}
\end{equation}
where $i=x,y,z$, $b_i(t)=R_i(t)/R_i(0)$ is the normalized radius and $V(t)= \prod_i b_i(t) $ is the normalized volume.
The $\bar{\gamma}$ is governed by the interaction strength $1/k_Fa$, where $k_F=\sqrt{2m(6N\times\omega_x\omega_y\omega_z)^{1/3}/\hbar}$
is the Fermi momentum and $a$ is the scattering length.
In the BEC limit ($1/k_Fa \gg1$) one has $\mu(r) \propto n(r)$ and thus $\bar{\gamma}=1$, while in the BCS limit ($1/k_Fa \ll-1$) and the unitary limit ($1/k_Fa=0$), $\mu(r) \propto n(r)^{2/3}$ leads to $\bar{\gamma}=2/3$.  Fig.~\ref{figure2} shows the experimental results of the aspect ratio $R_y(t)/R_x(t)$ at 767~G (BEC side), 832~G (unitary) and 1003.2~G (BCS side). Also shown are the numerical solutions of Eq.~(\ref{equ1})  (blue lines),
where the known $\bar{\gamma}$ values have been used.
Quantitative agreement is found during the early stage of expansion and up to a long time of over $11$~ms  on the BEC side and at unitarity;
the agreement is up to 6 ms on the BCS side, where fragile pairs might break during expansion.
Then, the general trends still agree qualitatively, while large deviations occur around the turning point. At the unitary and BEC side, we attribute these experiment-theory discrepancies to the limited imaging resolution of the experiment (as can be seen from $17$ ms image of Fig.~\ref{figure1}) and inadequate description of hydrodynamic equation around the turning point, where surface tension effect cannot be simply neglected. At the BCS side, the experiment-theory discrepancies are due to the breakdown of hydrodynamic description for the late-time expansion.

\begin{figure}[htbp]
  \centering
  \includegraphics[width=0.85\columnwidth]{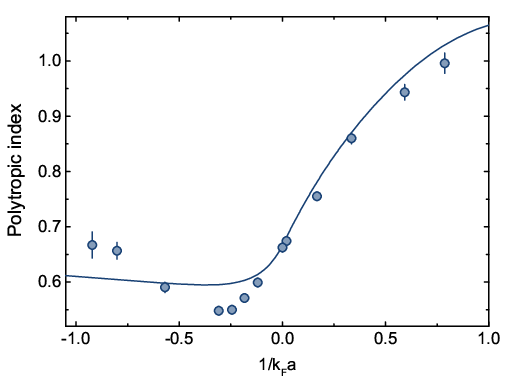}
  \caption{Polytropic index versus interaction strength. Blue circles are effective polytropic index $\bar{\gamma}$, while blue solid line is the numerical results of trap averaged polytropic index $\gamma'$. Error
bars represent one standard deviation. }\label{figure3}
\end{figure}

With the same procedure, the expansion dynamics of Fermionic superfluid is systematically studied in the whole BEC-BCS crossover and for various temperatures $T$. First, quantitative analyses are carried out to estimate the effective polytropic index  $\bar{\gamma}$ in the strongly interacting regime ($1/k_F|a| <1$), since it's the key parameter to model the hydrodynamic behavior of Fermionic superfluid. During expansion, the superfluid cloud  becomes less saturate and the measured radius $R_x(t)$ by absorption imaging is expected to be more precise. We fine tune the $\bar{\gamma}$ value in  Eq.~(\ref{equ1})
such that the numerical solution fits the experimental data with $t \leq 11$~ms for $1/k_Fa>-0.3$ and $ t \leq 6$~ms for $1/k_Fa\leq-0.3$.
The results of $\bar{\gamma}$ are shown in Fig.~\ref{figure3}, with  standard deviations less than 3.5\%.
The $\bar{\gamma}$ value smoothly decreases from the BEC to BCS side,  reaches a minimum value of $0.548(7)$ at $1/k_Fa \approx -0.3$,
and then gradually increases to the theoretical value $2/3$ in the BCS limit.

Although it's challenging to theoretically determine the $\bar{\gamma}$, the trap averaged polytropic index ${\gamma}'$ can be calculated thanks to the experimental progress on the measurement of EoS~\cite{Navon2010Nature,Ku2012Nature}. Using Gibbs–Duhem relation and EoS $\xi(\eta)$ in the canonical ensemble, the chemical potential can be expressed as $\mu(n)=E_F[\xi(\eta)-\frac{\eta}{5}\frac{\partial \xi(\eta)}{\partial \eta}]$, where $\eta=1/k_Fa$ is the interaction strength, $n$ is the density of atoms in single spin state,  and $E_F=\hbar^2k_F^2/2m$ is the Fermi energy, respectively.  Combine with $\mu(n)\propto n^\gamma$, the polytropic index $\gamma$ can be derived:
\begin{equation}\label{gamma}
\gamma(\eta)=\frac{n_f}{\mu}\frac{\partial \mu}{\partial n_f}=\frac{\frac{2}{3}\xi(\eta)-\frac{2\eta}{5}\xi'(\eta)+\frac{\eta^2}{15}\xi''(\eta)}{\xi(\eta)-\frac{\eta}{5}\xi'(\eta)}
\end{equation}
By substituting the experimentally extracted $\xi(\eta)$~\cite{Navon2010Nature} into Eq.~\ref{gamma}, the $\gamma(\eta)$ can be derived numerically. To further obtain the ${\gamma}'$, weight function $n(r_\alpha)r_\alpha^2$ ($\alpha=x,y,z$) of $\gamma(n)$ is introduced with $n(r_\alpha)$ being the density distribution of the Fermi gas in a harmonic trap. Then the ${\gamma}'$ can be numerically calculated by integrating over the trap as ${\gamma}'=\int d^3\mathbf{r} n(\mathbf{r})r_\alpha^2\gamma(n(\mathbf{r}))/\int d^3 \mathbf{r} n(\mathbf{r})r_\alpha^2$~\cite{HuiHu2004PRL}. The ${\gamma}'$ versus interaction strength is plotted in Fig.~\ref{figure3}. It's seen that ${\gamma}'$ is in good agreement with the $\bar{\gamma}$ in the unitarity and BEC side, and small deviation appears in the BCS side. We mention that, the polytropic index cannot be greater than 1 in general mean-field consideration. However, both the quantum Monte Carlo simulation and experimental measurement of EoS show that, due to strong correlations between the dimers, the polytropic index can exceed 1 in the strongly interacting regime at the BEC side.

\begin{figure}[htbp]
  \centering
  \includegraphics[width=\columnwidth]{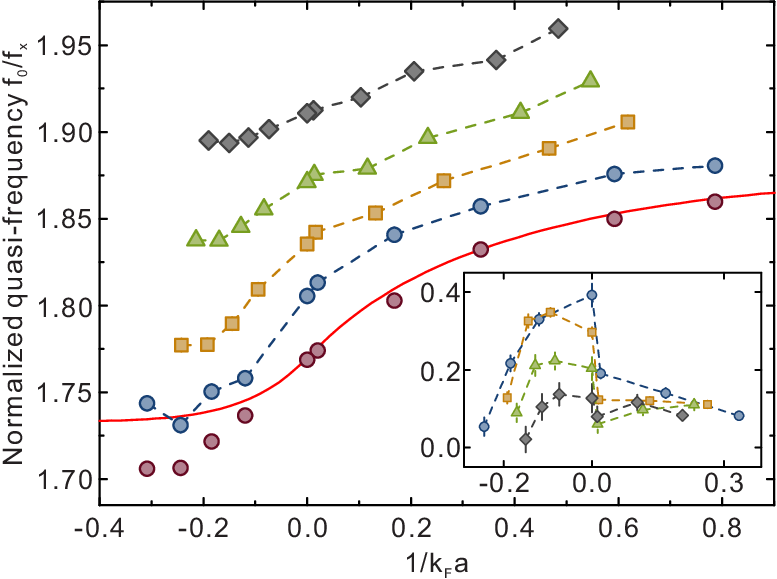}
  \caption{Normalized quasi-frequency $f_0/f_x$ versus interaction strength. Results represented by blue circle, yellow square, green triangle and black diamond are for $T/T_F$ of $0.06(1)$, $0.13(3)$, $0.18(2)$, and $0.27(4)$, respectively. The data points and error bars show the average and standard deviation of four measurements. Red circles and solid line are numerical solution of superfluid hydrodynamic equation using $\bar{\gamma}$ and ${\gamma}'$, respectively.  The inset shows  the absolute slope $|1/f_x\times\delta f_0/\delta (1/k_Fa)|$, as derived from the neighboring data points.}\label{figure4}
\end{figure}

To calculate the quasi-frequencies of the oscillatory expansion for different interaction strength, Eq.~\ref{equ1} is solved with the obtained $\bar{\gamma}$ and ${\gamma}'$, respectively. The results are shown in Fig.~\ref{figure4}, where the blue circles are the experimental results for the lowest temperature $T/T_F=0.06$, the red circles are the numerical results with $\overline{\gamma}$, and the red solid line represents the numerical results with $\gamma'$, respectively. The experimental $f_0/f_x$ drops significantly from the BEC side to the BCS side and reaches a minimum value of 1.73 around $1/k_Fa=-0.25$. It is observed that in the whole BEC-BCS crossover, the theoretical results of $f_0/f_x$ posses a smaller value than the experimental results. We attribute this deviation to the finite temperature of the atom cloud and possible pairing breaking during expansion, since Eq.~\ref{equ1} is valid for superfluid at zero temperature.

As the temperature goes up, the expansion of Fermi gas smoothly changes from superfluid hydrodynamics to classical hydrodynamics. The oscillatory expansions are still observed, with the decrease of normalized quasi-amplitude $A_0$ of the expansion decreases from $64\%\; (T/T_F = 0.06(1))$ to $25\%\; (T/T_F = 0.27(4))$. The theory-experiment gap increases, and the interaction strength dependence of normalized quasi-frequency $f_0/f_x$ are gradually smoothed out, mainly due to the decrease of superfluid component. With the further increase of temperature, the oscillatory behavior cannot be observed anymore, and the expansion will finally turn to a collisionless ballistic expansion. To further reveal the temperature and interaction effects, we calculate the absolute slope $|1/f_x\times\delta f_0/\delta (1/k_Fa)|$ from a pair of neighboring data points (see inset of Fig.~{\ref{figure4}}). The intriguing feature is a sudden increase appears near the resonance, particularly for the lowest temperature. It’s known that, while the BEC-BCS crossover is smooth on the macroscopic thermodynamic quantities, the crossover from cooper pairs to molecules will lead to a rapid change of the density and energy of the superfluid Fermi gas. A direct consequence is the fast change of the quasi-frequency with respect to the change of scattering length around the unitarity. For $T > 0.18(2) T_F$, the sudden increase is significantly suppressed, implying that the cloud might mostly consist of normal fluid, which is consistent with the superfluid phase transition temperature of $0.17 T_F$ at unitarity~\cite{Ku2012Nature}.

\section{Conclusion}

In summary, we successfully demonstrate that in the weak residual magnetic field curvature, an anisotropic Fermionic superfluid cloud experiences an oscillatory-like expansion for over 30 ms.
The oscillatory behavior is well supported by a superfluid hydrodynamic equation, and quantitative agreement is found during the early stage of expansion up to 11 ms. To sustain such a long time of expansion and oscillatory-like behavior, a large atom number of the atom cloud ($5\times10^6$ in this work) is needed. From the first several milliseconds of expansion, we extract for the whole BEC-BCS crossover the effective polytropic index, using which the quasi-frequencies  of the expansion are also calculated. The temperature effect on the quasi-frequency are further probed, as $T$ goes up, the interaction strength dependence of the quasi-frequency is smoothed out, implying the decreasing of superfluid fraction. We expect that with  large atom number and low temperature, high-precision study of many interesting static and dynamic phenomena in Fermionic superfluid becomes promising, which include elliptic flow~\cite{Cao2011NJP,Trenkwalder2011PRL}, normal-superfluid phase transition~\cite{HuiHu2010NJP}, and super-Efimov effect~\cite{Nishida2013PRL,ChaoGao2015PRA,Deng2016} etc.

\section*{A\lowercase{cknowledgment}}

This work was supported by NSFC of China (under Grants No. 11874340), the National Key R\&D Program of China (under Grants No. 2018YFA0306501), the CAS, the Anhui Initiative in Quantum Information Technologies and the Fundamental Research Funds for the Central Universities (under Grant No. WK2340000081).

\end{document}